# Energy dissipation and angular momentum transfer within a magnetically torqued accretion disc

GAN ZhaoMing, WANG DingXiong[†], LEI WeiHua & LI Yang

School of Physics, Huazhong University of Science and Technology, Wuhan 430074, China

**We discuss transportation and redistribution of energy and angular momentum in the magnetic connection (MC) process and Blandford-Payne (BP) process. The MC results in readjusting the interior viscous torque, and its effects are operative not only in but also beyond the MC region. The BP process is invoked to transfer the "excessive" angular momentum from an accretion disc. In addition, we derive a criterion for the interior viscous torque to resolve the puzzle of the overall equilibrium of angular momentum in disc accretion. It turns out that the BP efficiency of extracting angular momentum and the intensity of the outflow are required to be greater than some critical values.**

Disc accretion is widely regarded as an effective energy mechanism of high-energy phenomena in AGNs, X-ray binaries and some other compact objects. The interior viscous torque due to differential rotation transfers angular momentum outward to make accreting matter fall inward, and energy of disc radiation arises from dissipation of gravitational binding energy [1-4].

Rotating energy of a fast spinning black hole is another energy source for disc radiation. A black hole can exert a significant torque at its surrounding disc via large-scale magnetic field, and this mechanism is referred to as the magnetic connection (MC) process. The Poynting flux is injected into the inner disc in the MC process, resulting in the enhanced dissipation more concentrated in the inner disc [5, 6]. This feature can be used to interpret the very steep emissivity index of MCG-6-30-15 required by the observed Fe Kα line [7-10].

On the other hand, large-scale magnetic fields also play an important role in jet production, e.g. via Blandford-Payne (BP) process [11-14]. Recently, Miller et al. pointed out that the X-ray-absorbing wind discovered in the observation of GRO J1655-40 must be powered by a magnetic process, by which angular momentum can also be transferred efficiently out of the disc [15].

It is well known that angular momentum transfer is of fundamental importance in disc accretion. A potential barrier arising from continuous transfer of "excessive" angular momentum outward by the interior viscous torque blocks accretion at outer disc, and this problem is referred to as a puzzle of overall equilibrium of angular momentum of accretion discs. The following processes could be invoked to remove the "excessive" angular momentum from the disc:

(1) **Tidal interaction.** It has been argued that the tidal interaction with the companion star can exert a decelerating torque on the outer boundary, and transfer the excessive angular momentum of accretion disc to the bulk revolution of a binary system [16].

(2) **Radiation.** Angular momentum can be taken away from the disc by radiation. However, the efficiency is very low as argued in the next section of this paper. Probably, angular momentum can be carried away by gravitational wave in a very violent burst.

(3) **Exterior stresses**. Disc could be coupled to enwrapped corona, and the excessive angular momentum is transferred out of the disc by the stress exerted on the interface [17, 18]. However, the interface between the disc and corona is rather blurry, and the contribution due to the stress remains unclear.

(4) **Outflow.** Outflow can be launched from the accretion disc to carry away the excessive angular momentum. It is noted that the initial specific angular momentum of outflow matter is no greater than that of the accreting matter on the disc. Outflow can contribute to the overall equilibrium, provided that there exists some interaction between the outflow and the disc [19].

(5) **Large-scale magnetic field.** Analogous to the MC process, angular momentum can be taken away very efficiently by virtue of some "open" magnetic field lines anchored in the accretion disc in either Poynting or hydromagnetic fluxes, e.g., the BP process [11, 20-22].

In this paper, we will discuss the transportation and redistribution of energy in the MC process and the critical BP

Received; accepted
doi:
[†]Corresponding author (email: dxwang@mail.hust.edu.cn )
Supported by the National Natural Science Foundation of China under grant 10873005, the Research Fund for the Doctoral Program of Higher Education under grant 200804870050 and National Basic Research Program of China under grant 2009CB824800.

efficiency for the overall equilibrium of angular momentum. Throughout this paper the Boyer-Lindquist coordinates and the geometric units $G = c = 1$ are used.

## 1. Transportation and redistribution of energy in MC process

There are two kinds of large-scale magnetic field as shown in Figure 1. The MC process is invoked to transfer energy and angular momentum from a fast-spinning black hole to the inner region of accretion disc (henceforth, the MC region) via the closed field lines. The BP process is invoked to drive outflow centrifugally in the outer region of the disc (henceforth, the BP region). The boundary between the two regions is identified by $r_{tr}$, which can be determined by, e.g., the screw instability of magnetic field in the MC process [23].

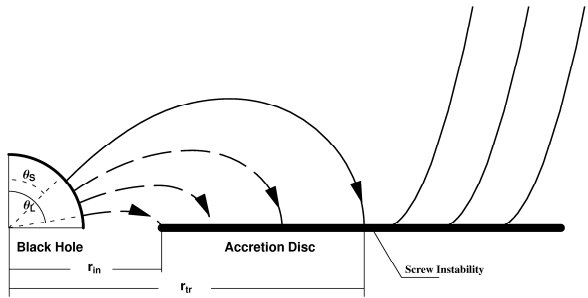

**Figure 1** Magnetic field configuration for the MC and BP processes.

The conservation equations of energy and angular momentum governing the disc accretion with the MC process are given as follows [5],

$$\frac{d}{dr}(\dot{M}_D L^\dagger - g) = 4\pi r(FL^\dagger - H_{MC}),  \quad (1)$$

$$\frac{d}{dr}(\dot{M}_D E^\dagger - g \cdot \Omega_D) = 4\pi r(FE^\dagger - H_{MC}\Omega_D),  \quad (2)$$

where $r$ is the disc radius, $\Omega_D$ is the angular velocity of the disc, and $\dot{M}_D$ is the accretion rate. The quantities $E^\dagger$ and $L^\dagger$ are the specific energy and specific angular momentum of the accreting matter, respectively [24]. The quantities $F$ and $g$ are respectively the radiation flux and interior viscous torque of the disc. The radiation flux consists of $F_{DA}$ and $F_{MC}$, which are the contribution due to disc accretion and MC process, respectively. The quantity $H_{MC}$ is the flux of angular momentum transferred from the black hole to the inner disc in the MC process, being related to the MC torque $T_{MC}$ as follows,

$$dT_{MC} = 4\pi rdr \cdot H_{MC},  \quad (3)$$

and the Poynting flux injected into the disc can be written as

$$S_{Poynting} = \frac{|\mathbf{E}_D \times \mathbf{B}_D|}{4\pi} = H_{MC}\Omega_D,  \quad (4)$$

i.e., the MC power transferred to unit disc area is exactly equal to the Poynting flux from the black hole to the accretion disc in the MC process, i.e., $p_{MC} = H_{MC} \cdot \Omega_D$ [10]. It is the magnetic torque that bends the field lines and induces a toroidal magnetic field, and subsequently produces the Poynting flux.

Contrastively, the accretion power arising from gravitational binding energy is expressed by the first term at LHS of eqn.(2), and we define the rate of releasing gravitational binding energy in unit disc area as

$$p_{DA} \equiv \frac{d(\dot{M}_D E^\dagger)}{4\pi r \cdot dr}.  \quad (5)$$

In this paper, we estimate the magnetic field on the black hole horizon by balancing the magnetic pressure and the ram pressure of accretion, and assume that the field on the disc varies with the disc radius in a power-law of index $n$ [10, 25]. The above equations are reduced to those for a standard thin disc by taking $H_{MC} = 0$.

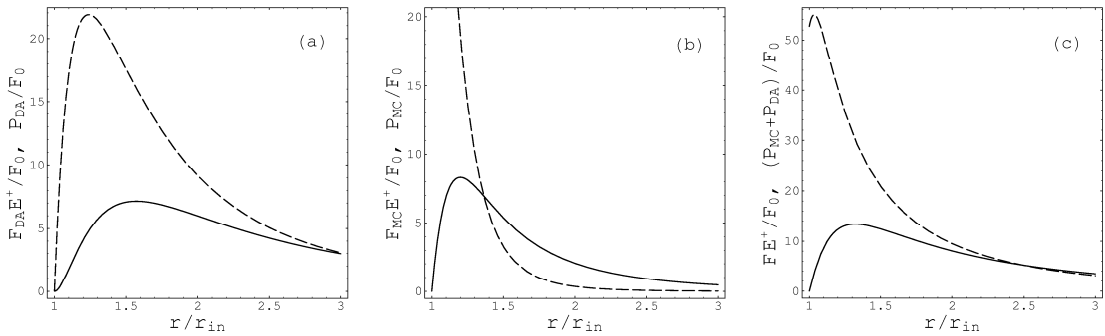

**Figure 2.** Comparison of different contribution to the radiation: (a) the accretion power $p_{DA}$ (dashed line) versus the locally dissipated energy flux $F_{DA} \cdot E^\dagger$ (solid line), (b) the MC power $p_{MC}$ (dashed line) versus the locally dissipated energy flux $F_{MC} \cdot E^\dagger$ (solid line), and (c) the total power $p_{DA} + p_{MC}$ (dashed line) versus the total locally dissipated energy flux $F \cdot E^\dagger$ (solid line) with black-hole spin $a_* = 0.5$ and $n = 6$.

Inspecting eqn.(1) and (2), we find that the MC effect on the disc radiation is validated by readjusting the interior viscous torque. Moreover, as shown in Figure 2 the total released energy at a disc radius, including energy released from gravitational binding energy and that transferred in the MC process, cannot be dissipated locally because of the conductive effect of the interior viscous torque notated by the second term at LHS of eqn.(2). Therefore a part of the released energy is transferred from the inner disc to the outer disc, and the MC effect works not only in the MC region but also in the BP region.

It should be emphasized that the viscous torque only transfers the angular momentum outward within the disc, and it has no contribution to the overall equilibrium. The disc radiation is very ineffective in transferring angular momentum, and it takes away only a very small fraction of the total "excessive" angular momentum as shown in Figure 3.

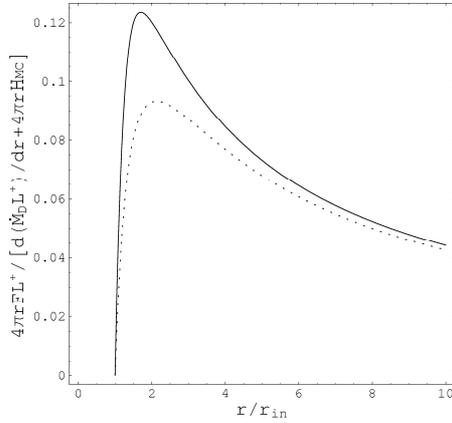

**Figure 3** Fraction of total "excessive" angular momentum taken away by radiation with and without MC effects in solid and dotted lines, respectively. The parameter $a_*$ and $n$ are the same as in Figure 2.

Thus tremendous angular momentum is accumulated persistently at the outer disc, i.e.,

$$\frac{dg}{dr} > 0 \quad \text{and} \quad g(r \to \infty) \to \infty, \quad (6)$$

How to carry the excessive angular momentum out of accretion disc to guarantee a steady disc accretion? It is an open question to a consistent theory of accretion disc [18].

## 2. Critical BP efficiency for overall equilibrium of angular momentum

Since the excessive angular momentum cannot be taken away efficiently via radiation, outflow is invoked to transfer angular momentum from the disc. It turns out that the puzzle of overall equilibrium of angular momentum can be resolved, provided that the efficiency of extracting angular momentum via the BP process and the outflow rate are greater than some critical values.

In our model energy is extracted from disc in the form of Poynting flux because of the magnetic interaction, and the latter is converted into the kinetic energy of the matter flowing along the field lines. As shown in ref.[11], the specific energy $e$ and the specific angular momentum $l$ of the outflow along a field line remain constant, where

$$l = r_{jet}\upsilon_\varphi - r_{jet}B_\varphi/k. \quad (7)$$

The quantities at RHS of eqn.(7) have the same meanings as given in ref.[11], and $l$ is related to the specific angular momentum of accreting matter as follows,

$$l = \lambda L^\dagger, \quad (8)$$

where the parameter $\lambda$ represents the efficiency of extracting angular momentum from the disc. Based on the conservation of angular momentum we have the following relation in the BP region,

$$\frac{d}{dr}(\dot{M}_D L^\dagger - g) = 4\pi r F L^\dagger + l \cdot \frac{d\dot{M}_D}{dr}. \quad (9)$$

Eqn.(6) implies that some mechanisms are required to remove angular momentum away from the disc. In addition, the efficiency of the mechanism should be high enough to make the viscous torque decrease with the disc radius, attaining a small value eventually at the outer boundary of the disc. To get the critical BP efficiency for the overall equilibrium of angular momentum, we set

$$dg/dr < 0. \quad (10)$$

Incorporating eqns.(8)-(10), and neglecting the contribution of disc radiation to the angular momentum transfer, we have

$$\frac{d\ln L^\dagger}{dr} < (\lambda - 1)\frac{d\ln \dot{M}_D}{dr} \quad (11)$$

Since the BP region is far from the hole ($r_{tr} > r_{in} > r_H$), we take Newtonian expression for the specific angular momentum, i.e., $L^\dagger \approx \sqrt{GMr}$. Considering the existence of the outflow, we assume the accretion rate varying with the disc radius as follows,

$$\frac{d\ln \dot{M}_D}{dr} = \frac{s}{r}, \quad r > r_{tr}, \quad (12)$$

where the parameter $s$ ($0 < s < 1$) represents the intensity of outflow. Incorporating eqns.(11) and (12), we have

$$2s(\lambda - 1) > 1. \quad (13)$$

Eqn.(13) implies that the viscous torque could decrease with the disc radius, provided that the efficiency of extracting angular momentum by the BP process (represented by $\lambda$) and the intensity of the outflow (represented by $s$) are high enough. This is a critical condition for the overall equilibrium of angular momentum in accretion discs.

## 3. Conclusion and discussion

In this paper, we discuss the MC and BP effects on transportation and redistribution of energy and angular momentum in accretion discs. It turns out that the MC effect readjusts the interior viscous torque, which is valid not only in the MC region but also in the beyond BP region. We argue that the BP process is of essential importance for the overall equilibrium of angular momentum in accretion discs, and a criterion is derived in terms of the BP efficiency of extracting angular momentum and the intensity of the outflow. This result implies that disc accretion and outflow/jet are essentially symbiotic and inseparable.

According to the criterion (13) the BP efficiency of extracting angular momentum must be larger than unity, i.e., $\lambda > 1$, and this implies the presence of some interaction between disc and outflow. In fact, a negative work is always done by the magnetic torque due to the presence of large-scale poloidal magnetic field in the BP process, acting at the induce current flowing radially on the disc. It is very similar to the case that an external magnetic field always exerts torque on the induce eddy currents in a conductive disc. The parameter $\lambda$ and the outflow rate $s$ are not independent. Essentially, $\lambda$ is related to the accelerating process of outflow, and it depends not only on the intensity and shape of large-scale magnetic fields, but also on the outflow rate. The outflow with greater $s$ leads to less $\lambda$, since the outflow matter cannot be constrained rigidly to the field lines due to its inertia [14]. The detailed process of accelerating outflow is very complex, which is beyond the scope of this paper.